\documentclass[sigconf]{acmart}
\AtBeginDocument{%
  }

\setcopyright{acmlicensed}
\copyrightyear{2026}
\acmYear{2026}
\acmDOI{XXXXXXX.XXXXXXX}
\acmISBN{978-1-4503-XXXX-X/2018/06}
\newcommand{\nop}[1]{}

\definecolor{Imine}{rgb}{0.0, 0.2, 1}
\definecolor{Xiang}{rgb}{1,0,1}
\definecolor{Kambiz}{rgb}{0.5,0.1,1.0}

\usepackage{balance}
\usepackage{enumitem}
\renewcommand{\shortauthors}{}

\begin{document}
\sloppy

\title{Towards a Privacy-Preserving Reliability-Aware Entity Resolution for Attributed Graphs [Vision] (Technical Report)}

\author{Abdessamad Imine}
\affiliation{%
\institution{Universit\'e de Lorraine, CNRS, Inria, LORIA}
\city{54000 Nancy}
  \country{France}
}
\email{abdessamad.imine@loria.fr}

\author{Xiang Lian}
\affiliation{
\institution{Department of Computer Science, Kent State University}
\city{Kent, OH}
  \country{USA}
}
\email{xlian@kent.edu}

\author{Kambiz Ghazinour}
\affiliation{
\institution{Department of Cybersecurity, State University of New York at Canton}
\city{Canton, NY}
  \country{USA}
}
\email{ghazinourk@canton.edu}

\begin{abstract}

\textit{Entity resolution} (ER) on attributed graphs aims to identify nodes across multiple graphs that correspond to the same real-world entity and merge them into an integrated graph. When input graphs vary in quality and contain sensitive information, ER faces a dual challenge: while reliability information is crucial for accurate graph data integration, its presence in the published integrated graph can heighten privacy risks. This paper presents a research vision that unifies reliability assessment and privacy preservation within a single graph ER process. We highlight the inherent tension between these two objectives, propose a three-level coupled framework, and outline the remaining research challenges to enable the publication of integrated graphs that are both accurate and privacy-preserving.

\end{abstract}


\maketitle

\section{Introduction}

\textit{Entity resolution} (ER) is a fundamental task in the fields of data integration and knowledge fusion \cite{bhattacharya2006entity}. When input data are represented as attributed graphs, where nodes carry attributes and edges encode relationships, ER extends beyond simple record matching to encompass structural and semantic alignment between different graphs. 
For instance, in the healthcare sector, clinical care teams (doctors and nurses) act as key end users, leveraging the resolved graph to view an error-free, unified patient history that prevents dangerous medication or treatment mistakes. Additionally, hospital compliance officers and billing administrators utilize this clean graph structure to detect insurance fraud, eliminate duplicate medical records, and track regional public health trends accurately.

In practice, the input graphs $G_1$, $G_2$, $\ldots$, and $G_n$ originate from heterogeneous sources with varying levels of reliability: government databases may be considered highly reliable, web-scraped data are often noisy, while crowdsourced annotations generally fall somewhere between these two extremes.
A natural approach is to leverage reliability information to guide ER decisions,  with the aim of producing an integrated graph. Items from highly reliable sources should be given greater weight, whereas information from less reliable sources should be treated with caution. Keeping reliability information within the integrated graph is important, namely:

\noindent (i) {\em Downstream tasks require reliability}: if we subsequently perform community detection, link prediction, or graph analysis, knowing which edges or nodes are reliable helps avoid drawing conclusions with overconfidence.

\noindent (ii) {\em Uncertainty propagation}: When querying the graph (for example, to ``find paths between $N$ and $M$''), unreliable edges
should contribute less to the confidence level of the answer.

\noindent (iii) {\em Transparency}: Users and analysts need to know which parts of the integrated graph come from high-reliability sources
versus low-reliability sources.

However, incorporating reliability into ER processes/results comes at a cost. When the integrated graph is published, the reliability metadata it contains can serve as a roadmap for attackers: edges and attributes marked as highly reliable directly indicate which information constitutes the most attractive targets. Research into privacy-preserving graph publication has shown that node identity disclosure and link re-identification are two major attack objectives \cite{Yang2023}. Reliability information can amplify both risks.

\begin{figure*}
    \centering
    \includegraphics[width=1.05\linewidth]{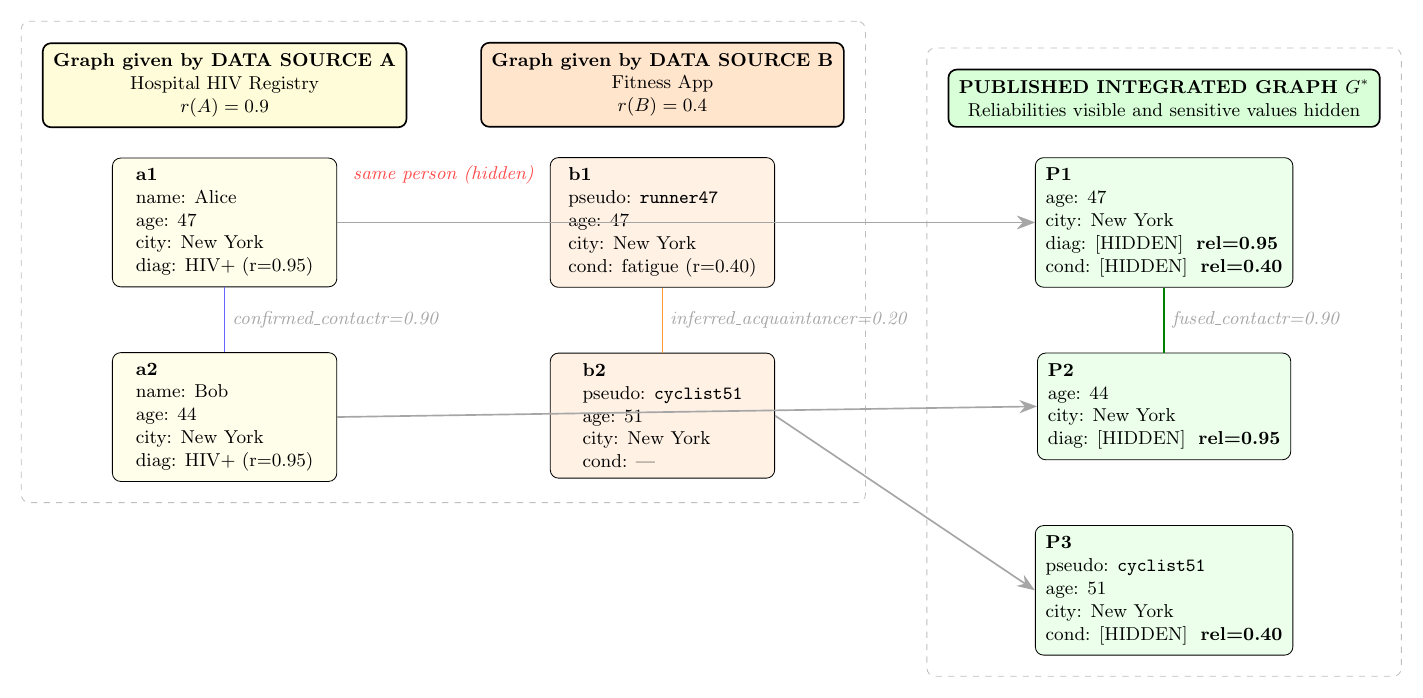}
    \caption{An illustrative example on how privacy leaks happen (Example 1.1).}
    \label{fig:Figure1}
\end{figure*}

To illustrate this privacy risk we provide the following example:
\begin{example}\label{Example1}\it
A healthcare institution collects graphs from two data sources, computes their reliability (based on some computation method), and performs ER to produce an integrated graph (see Figure \ref{fig:Figure1}). Source $A$ is a hospital HIV clinic registry with reliability $r(A)=0.9$: its nodes carry lab-confirmed diagnoses, and its edges encode confirmed transmission contacts. Source $B$ is a fitness tracking application with reliability $r(B)=0.4$: its nodes carry self-reported conditions, and its edges encode inferred acquaintances.
Node $a_1$ in Source $A$ represents Alice ($47$, New York, HIV+) and is connected to $a_2$, Bob ($44$, New York, HIV+) by a confirmed-contact edge with reliability $0.90$. Node $b_1$ in Source $B$ represents the same Alice under the pseudonym \verb+runner47+ (self-reported fatigue), connected to $b_2$, \verb+cyclist51+ ($51$, New York) by an inferred-acquaintance edge with reliability $0.20$.
The ER process recognizes that $a_1$ and $b_1$ refer to the same person and merges them into a single node $P_1$; it merges $a_2$ into $P_2$ and keeps $b_2$ as a separate node $P_3$. During fusion, conflicting attribute values are resolved in favor of the more reliable source, so $P_1$'s diagnosis inherits reliability $0.95$ from Source $A$, while $P_1$'s self-reported condition inherits reliability $0.40$ from Source $B$; the edge ($P1$,$P_2$) inherits reliability $0.90$, dominated by Source $A$'s confirmed contact.


The integrated graph is published $G^*$ with sensitive attribute values hidden but all reliability scores visible, believing that hiding the values is sufficient for privacy. This belief is false. Because the fused reliability of an attribute is a function of which sources contributed, and because the two sources have very different reliabilities ($0.9$ versus $0.4$), the published reliability score makes it possible to directly identify the source from which it originates. 
An attacker knows the two sources, their reliabilities, the fusion rule (e.g., conflicting values are resolved in favor of the more reliable source) and Alice is $47$, lives in New York, and uses the fitness application under the pseudonym \verb+runner47+, but they do know whether Alice's hospital record exists.
The attacker observes $P_1$'s diagnosis reliability of $0.95$ and reasons that only the hospital registry could produce such a high value;
Since the hospital is specifically an HIV clinic, the attacker concludes that Alice has HIV.
Similarly, the observed edge reliability of $0.90$ on ($P_1$,$P_2$) can only arise from Source $A$'s confirmed-contact edge, revealing that Alice is in a documented HIV transmission cluster with another patient (namely Bob). By contrast, the observed reliability of $0.40$ on $P_3$'s condition tells the attacker that only the fitness app contributed, so \verb+cyclist51+  is not in the hospital registry. The reliability scores thus serve as a covert channel that discloses HIV status, cluster membership, and source membership (without any sensitive attribute value ever being published) which precisely demonstrates that anonymization alone is insufficient. \quad $\blacksquare$%
\end{example}

This paper proposes a research vision: \textit{reliability computation} and \textit{privacy preservation} should be designed jointly within the ER pipeline, not treated as independent post-processing steps. Our goal is not to ``compute reliability first, then add privacy'', but to incorporate privacy constraints into reliability estimation, allocate privacy budgets in merge decisions, and release an integrated graph that achieves a controllable balance between accuracy and privacy.

The rest of the paper is organized as follows.
Section 2 elaborates on the problem statement that the authors are interested in. Section 3 discusses the reliability computation for edge and attribute components. Section 4, talks about interaction between reliability-aware entity resolution and privacy preservation. Section 5 invites the readers to the research vision of the paper and shares some open challenges with the interested audience. Section 6 discusses some great work in the literature and Section 7 concludes the paper.
\section{Problem Statement}\label{sec:ps}

A {\em trusted third party} collects $n$ undirected, attributed graphs $G_1$, $G_2$, $\ldots$, and $G_n$, each from a data source, and aims to perform an ER operation to produce an integrated graph $G^*$ in which each node corresponds to a real-world entity and each edge to a real-world relationship. Two problems should be addressed:


\vspace{1ex}\noindent\textbf{Problem 1: Reliability computation.}  Before ER, compute reliability scores for edges and attributes in each $G_i$. These scores must reflect fine-grained trustworthiness at the edge and attribute levels.

\vspace{1ex}\noindent\textbf{Problem 2: Privacy preservation.}  When publishing $G^*$, ensure that node identities and sensitive links cannot be inferred by attackers using auxiliary information.

The reliability computation is highly  complex due to the exponential growth in the number of possible combinations of nodes and edges. Furthermore, the presence of overlapping paths and cyclic loops renders these combinations statistically dependent, making exact computations mathematically overwhelming for real-world graphs \cite{Fan22}.

Reliability plays a dual role in ER. First, as {\em a signal}, it distinguishes reliable evidence from noise, thereby making merging decisions more precise. Secondly, it represents {\em a risk}, as it highlights the most certain and sensitive parts of the integrated graph, thereby offering high-value targets to attackers (as illustrated in Example \ref{Example1}). This tension has received insufficient attention in the field of privacy-preserving graph publication.


Differential privacy (DP) has emerged as the gold standard for graph publication because it offers rigorous privacy guarantees that are independent of the attacker,  whereas $k$-anonymity and other methods lacking formal guarantees rely on assumptions about the attacker's background knowledge that fails as that knowledge grows \cite{dwork2014,Sok26}. While cryptography ensures robust security for specific computations, it does not provide a formal privacy guarantee against inferences drawn from published results, making DP a more suitable foundation for publishing a graph intended for unknown adversaries.

Existing methods for differentially private graph publication focus primarily on structural perturbation and query utility, yet rarely consider the impact of reliability or confidence metadata on privacy semantics \cite{Yang2023}. Similarly, developing a theoretical framework for privacy-preserving ER over graphs remains an open problem.


\section{Reliability Computation: From Source-Level to Evidence-Level}\label{sec:reliability}
Reliability is not merely a single value associated with a graph. It is a fine-grained property of the evidence that the graph carries, and it must be computed at the level of individual edges and attribute values before it can guide ER.

\subsection{Multi-Source Estimation of Edge Reliability}
\label{subsec:edge_rel}

For an edge $e=(u,v)$ in graph $G_i$, its reliability $r_{edge}(e)$ must be estimated along at least three dimensions:
\begin{enumerate}[leftmargin=*]
\item {\em Structural signals.} Triadic closure serves as a reliable indicator of the credibility of an edge in social networks: if two nodes share many common neighbors, the edge is more likely to represent a real relationship than random noise. Edge betweenness offers complementary insight: edges with low betweenness generally correspond to core relationships, whereas those with high betweenness may act as bridges between communities and require more careful handling in the context of ER.


\item {\em Source metadata.} If we can compute a global reliability $r(G_i)$, a preliminary edge reliability for  $e=(u,v)$ in $G_i$ can be estimated as $r(G_i)\cdot c(e)$, where $c(e)$ is edge-level confidence (e.g., sensor signal strength, extraction model confidence, etc.).

\item {\em Cross-graph consistency.} Following an alignment of certain nodes via ER, the frequency of edges between the graphs can be calculated. This method carries a risk of circularity: alignment errors skew reliability estimates.
\end{enumerate}

For the three dimensions of edge reliabilities above, we can devise an edge aggregation function $f_{edge}(\cdot)$ (e.g., weighted sum) to combine them and obtain an edge reliability score $r_{edge}(e) \in [0, 1]$.

\subsection{Fine-Grained Decomposition of Attribute Reliability}
\label{subsec:attr_rel}

The reliability of an attribute must not be a global constant. For attribute $a$ of node $v$ with value $x$, reliability is broken down into:

\begin{enumerate}[leftmargin=*]
\item {\em Global attribute reliability.} We denote by $r_{attr}(G_i, a)$ the overall collection quality of attribute $a$ in graph $G_i$. For example, manually entered ``name'' might have reliability around $0.9$, while OCR-extracted ``address'' might be around $0.7$.

\item {\em Value-level reliability.} We denote by $r_{val}(v, a, x)$ the trustworthiness of this specific value. This estimate can be derived from the frequency of values within the domain, format compliance, and consistency between attributes. An unusual name spelling may warrant a lower weighting, whereas a correctly formatted email address inspires greater confidence.

\item {\em Domain constraint reliability.} We also need the definition of the attribute domain itself to be reliable. For example, is the value range $[0, 120]$ for the ``age'' attribute strictly enforced? Thus, we use $r_{cons}(v, a)$ to reflect the logical consistency of this attribute value with other attributes.
\end{enumerate}

A general composite formula of attribute reliability $r_{attr}(v,a) \in [0, 1]$ is given as follows:
\[
r_{attr}(v,a) = f_{attr}(r_{attr}(G_i, a), r_{val}(v,a,x), r_{cons}(v,a)),
\]

\noindent where $f(\cdot)$ is an attribute aggregation function of three types of attribute-related reliability levels. 

One interesting research direction is to design appropriate attribute aggregation function $f_{attr}(\cdot)$ (e.g., multiplication or weighted sum) to mimic accurate attribute reliability $r_{attr}(v,a)$.

\subsection{The Reliability Inference} 

\vspace{1ex}\noindent {\bf Challenges.} Given graphs $G_i$ from multiple data sources, it is important, yet challenging, to infer the reliability, $r(G_i)$, of graphs $G_i$ (i.e., data sources) and that, $r_{edge}(e)$ or $r_{attr}(v, a)$, of edges/attributes in $G_i$. First, although we can use an aggregation function to obtain reliability scores (as discussed in Sections \ref{subsec:edge_rel} and \ref{subsec:attr_rel}) correlated with (i.e., proportional to) the probabilities that graphs/edges/attributes are true, there still exists a gap between reliability scores and semantically meaningful reliability probabilities. Moreover, due to large scale of graphs $G_i$, it is also non-trivial how to efficiently infer such reliability probabilities. 

\vspace{1ex}\noindent {\bf Potential Research Direction.} One important research direction is on accurate and efficient inference of reliability probabilities of graph components (e.g., edges, node attributes, and graphs) in a scalable manner.



{\it Accurate Reliability Inference:} 
We initialize reliability probability, $r(c)$, of each edge $e$ (or attribute $a$) with reliability score $r_{edge}(e)$ (or $r_{attr}(v, a)$). Then, the reliability, $r(G_i)$, of each graph $G_i$ (data source) is: $$r(G_i) = \frac{\sum\limits_{c \in C(G_i)} r(c)}{|C(G_i)|},$$ where $C(G_i)$ is a set of graph components such as edges $e$ and/or attributes $a$ in graph $G_i$, and $|\cdot|$ is the size of a set.


Then, we can also compute the edge/attribute reliability value $r(c)$, that is: $$r(c) = 1 - \prod\limits_{\forall G_i, c \in C(G_i)} (1 - r(G_i)),$$ which is given by the probability that this attribute value is true in graph(s) $G_i$ from at least one data source. 


We iteratively utilize the two equations above to compute the reliability, $r(G_i)$, of graphs $G_i$, which are used to calculate edge/attribute reliability $r(c)$ again. This way, we can obtain both graph reliability, $r(G_i)$, and edge/attribute reliability, $r(c)$, until these reliability scores converge. The converging scores are then considered as reliability probabilities.

{\it Efficient Reliability Inference: } To improve the efficiency and scalability of reliability inference over large-scale graphs, database techniques such as pruning and indexing approaches can be used to speed up the inference process. As an example, we may group graph components (e.g., edges or node attributes) with similar reliability and distributions among data sources as clusters, on which we can build indexes for quick accesses (via pruning) and updates. 

\begin{figure*}[t!]
\centering 
\includegraphics[width=0.7\textwidth]{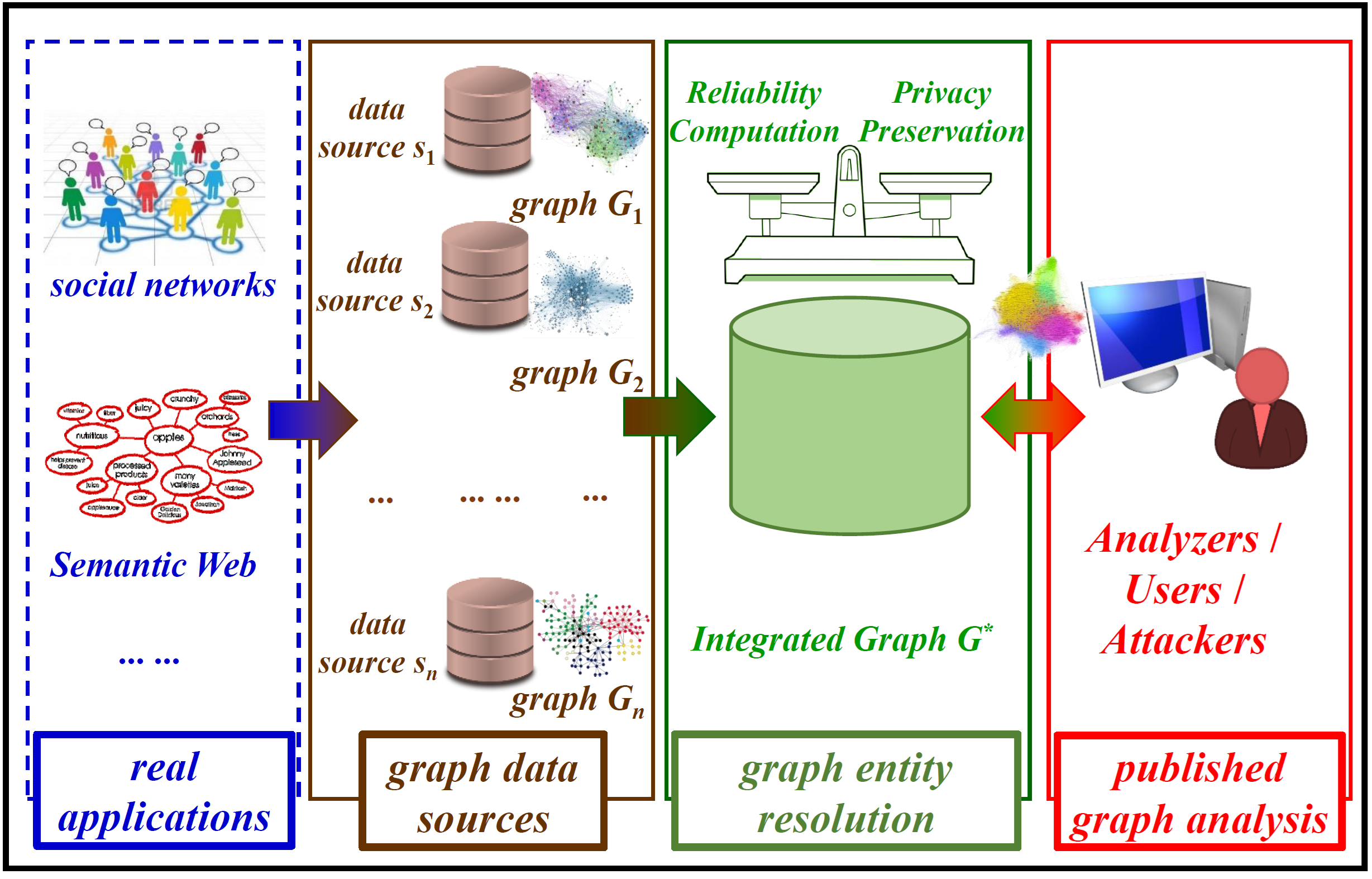}
\caption{\small A unified graph ER framework for privacy-preserving reliability-aware
entity resolution.}
\label{fig:framework}
\end{figure*}

\section{Interaction Between Reliability-Aware ER and Privacy Preservation}
\label{sec:privacy}

Having defined how reliability is computed, we now examine what happens when the resulting metadata is published: reliability is not a neutral annotation, but a source of risk through which hidden information regarding input data sources can reach an attacker.



\subsection{How Reliability Information Changes Privacy Computation}
When the integrated graph includes reliability information, the quantification of privacy-related risks must be extended. The traditional risk of node identity disclosure considers only structural characteristics \cite{Yang2023}. A version incorporating reliability must also take into account:

\vspace{1ex}\noindent{\bf Anonymization is not sufficient.}
Traditional anonymization removes direct identifiers and may generalize quasi-identifiers, but it does not alter reliability annotations. Since the reliability channel depends on the annotations rather than the attribute values, anonymization leaves this channel intact. In Example \ref{Example1}, masking the value "HIV+" is futile if a reliability score of $0.95$ still signals to an attacker that a hospital diagnosis contributed to the result. The channel persists because it is conveyed by a derived statistic whose value depends on the hidden source configuration, not by the value that was hidden.

\vspace{1ex}\noindent{\bf Reliability as enhanced structural fingerprint.} A node's reliability score vector (i.e., the combination of its reliability value across attributes) constitutes a high-dimensional fingerprint in itself. Attackers can exploit these scores to perform more precise node matching. For example, given two input graphs $G_1$ and $G_2$, if an attacker knows that the target exhibits high reliability in $G_1$ but low reliability in $G_2$, nodes in the integrated graph corresponding to this reliability profile will be prioritized.

\vspace{1ex}\noindent{\bf Reliability as a premise for link inference.} High-reliability edges are explicitly identified in the integrated graph, allowing attackers to prioritize them for the extraction of sensitive information. If a sensitive edge has a high reliability score, the attacker can reasonably infer that the relationship is real, thereby successfully re-identifying the link.


\subsection{Differential Privacy (DP) as the Solution}

We now examine how reliability computation and privacy preservation interact in the ER process. Our objective is to identify precisely why reliability metadata is a source of privacy risk, and to state the privacy requirement that any reliability-aware ER mechanism must satisfy.

\vspace{1ex}\noindent{\bf Attacker model.} The attacker knows the input graph schema, the overall reliability of each source, the ER algorithm, the reliability computation, the privacy mechanism, and the records of all individuals except the target $T$. The attacker possesses auxiliary knowledge $K_T$ regarding $T$ (e.g., quasi-identifiers such as age, city, or known source membership), but is unaware of whether records from $T$ are present in the input graphs and which sources contain them. This represents the standard worst-case attacker for DP, adapted to the context of ER.

\vspace{1ex}\noindent{\bf Reliability as a risk source.}
Let $G_T \subseteq \{G_1, \ldots, G_n\}$ be the (unknown) set of input graphs containing $T$'s records. The fused reliability of an attribute or an edge in the published graph depends on the reliability of the sources that contributed to this fusion. Since source reliabilities vary, the published reliability score is a function of $G_T$. This dependency makes thus reliability as a risk source: it represents the conditional distribution of the published reliability annotations, given the hidden configuration $G_T$ and the attacker's auxiliary knowledge $K_T$. This risk source is significant because fusing information from different sources can yield distinct reliability values. In Example \ref{Example1}, a fused diagnostic reliability of $0.95$ can be obtained only if the hospital source contributed, whereas a value of $0.40$ indicates that only the fitness application contributed. The published reliability score thus reveals the source, and this source information is precisely what the attacker seeks to infer.

\vspace{1ex}\noindent{\bf Extension of DP semantics.}
The privacy requirement for a reliability-aware ER mechanism is {\em reliability-aware DP}.
Let $\mathcal{C}=(G_1, \ldots, G_n)$ be the tuple of inputs graphs and $\mathcal{C}_{-T}$ the tuple obtained by removing all $T$'s records.
A randomized mechanism $\mathcal{M}$ satisfies ($\epsilon$,$\delta$)-reliability-aware DP if, for every target $T$, every pair ($\mathcal{C}$,$\mathcal{C}_{-T}$), every auxiliary knowledge $K_T$, and every measurable set $S$ of published outputs:
\[
\text{Pr}[\mathcal{C}\in S\,\vert\,  K_T] \leq e^\epsilon \text{Pr}[\mathcal{C}_{-T}\in S \,\vert\, K_T]+\delta
\]
The set $S$ ranges over subsets of the full published output, including the structure, the attribute values, and the reliability annotations jointly. This joint requirement is essential: a mechanism that protects structure and values but publishes unperturbed reliabilities does not satisfy the guarantee, because the reliabilities leak information about the hidden configuration.
According to this definition of reliability-aware DP, the amount of information carried by the reliability metadata is bounded by a quantity that depends only on $\epsilon$ and $\delta$, independently of how informative the reliability scores would otherwise be.

\subsection{A Key Design Principle}
From the analysis above, we derive a fundamental design principle:

{\em The privacy cost associated with reliability information is proportional to its informational content. The more precise and detailed this information is, the more valuable it is to attackers, and, consequently, the higher the level of privacy protection required to offset this risk.
}

This principle implies that reliability computation and privacy preservation cannot be designed independently. As early as the reliability estimation stage, the following factors must be considered: (i) the level of granularity at which this reliability score will appear in the published graph, (ii) whether it requires perturbation, and (iii) from which privacy budget this perturbation is drawn.

\section{Research Vision and Open Challenges}
\label{sec:challenges}

The tension between reliability and privacy is not a flaw to be eliminated, but a structural feature of the problem and addressing it requires a research program rather than a single technique.

\subsection{A Vision for a Unified Framework}
We envision a unified graph ER framework as illustrated in Figure \ref{fig:framework}, which collects graphs $G_1$, $G_2$, ..., and $G_n$ from data sources $s_1$, $s_2$, ..., and $s_n$, respectively, by a trusted third party, performs graph entity resolution among these $n$ graphs, taking into account both aspects of reliability and privacy preserving, and publishes the integrated graph $G^*$ for analyzers, users, or even attackers to analyze. 

Specifically, the reliability computation and privacy preservation in this framework are coupled at three layers:

\vspace{1ex}\noindent{\bf Layer 1: Privacy constraints related to reliability estimation.} When calculating the reliability of edges and attributes intended for publication, the estimation process itself must satisfy DP requirements. For example, calculating triadic closure involves counting queries over sets of neighbors. It is possible to design approximate counting mechanisms that guarantee ($\epsilon$,$\delta$)-DP.

\vspace{1ex}\noindent{\bf Layer 2: Privacy budget allocation for merge decisions.} Matching decisions in ER (i.e. determining whether a pair of nodes should be merged) are inherently sensitive: they reveal which entities correspond to the same real-world object. In privacy-preserving ER, these decisions must be governed by a privacy budget. Reliability information can guide the allocation of this budget: decisions supported by high-reliability evidence can be made at a lower privacy cost, because even if the decision is leaked, the attacker gains less information.

\vspace{1ex}\noindent{\bf Layer 3: Perturbation of reliability in the published graph.} 
When publishing the integrated graph, reliability scores must be perturbed using a DP mechanism. Rather than publishing exact values, reliability can be discretized into a limited number of categories (e.g., "high/medium/low"), with category assignments published under the ($\epsilon$,$\delta$)-DP model. DP is crucial here: unlike the generalization used in $k$-anonymity, it provides a formal, adversary-independent guarantee that bounds the amount of information conveyed by the published scores. Furthermore, it allows for the composition of guarantees across the various stages of the ER pipeline (reliability computation, merging, and publication). The categorization step reduces the sensitivity of the published statistic, thereby enabling more efficient use of the privacy budget, while the DP mechanism ensures that residual information regarding a specific individual's data remains bounded by $\epsilon$ and $\delta$.

\subsection{Open Research Questions}

We now list the research questions that must be answered for this vision to become a reality.

\vspace{1ex}\noindent{\bf Question 1: Reliability-aware privacy semantics.} 
This involves defining ``reliability-aware DP'' and extending the notion of neighboring graphs to a ``reliability-based neighborhood''. Additionally, it requires analyzing how the parameters $(\epsilon, \delta)$  vary according to the granularity of reliability.

\vspace{1ex}\noindent{\bf Question 2: Privacy-utility trade-off in reliability computation.} 
Reliability estimation exploits internal graph structure, such as triadic closure and neighbor consistency. Under DP, these statistics require noise injection, which degrades the quality of the resulting reliability scores. A key open problem is to characterize how this noise propagates through the reliability computation and to determine whether there exists a lower bound on reliability estimation precision below which the reliability information loses its value for guiding ER.

\vspace{1ex}\noindent{\bf Question 3: Defenses against reliability fingerprinting.} If reliability score vectors constitute node fingerprints, how can defenses be designed? Possible approaches include randomization of reliability scores, cross-node reliability smoothing, and aggregate publication of reliability information (e.g., publishing global reliability distributions rather than node-level reliabilities).

\vspace{1ex}\noindent{\bf Question 4: Dynamic data scenarios.} When input graphs are updated over time, reliability scores and privacy protection require incremental updates. Continuous publication under DP faces the problem of privacy budget exhaustion. The dynamic publication of reliability information further complicates the issue: fluctuations in reliability can themselves lead to information leakage.

\vspace{1ex}\noindent{\bf Question 5: Evaluation methodology.} How can an integrated graph that is both ``accurate and privacy-preserving'' be evaluated? Standard ER metrics (precision, recall, F1-score) and privacy metrics (privacy loss, attack success rate) must be integrated into a unified evaluation framework. In particular, how does the inclusion of reliability information alter the utility-privacy trade-off curve?

\vspace{1ex}\noindent{\bf Question 6: Personalized privacy preferences.} Considering that in reality, each data provider might have different privacy preferences on how each of their attributes are handled, there needs to be a framework in which each data provider's preferences towards each attribute are captured. In other words, there needs to be an additional value for each data item that is defined by the data provider as the sensitivity value of the data item. An item might be highly sensitive for one individual, and not sensitive for another one. These should reflect each data provider's privacy preferences.

\section{Related Work}

We position our vision relative to existing work: show how our vision builds upon, but is distinct from, current research trends.


\vspace{1ex}\noindent{\bf Entity Resolution (ER).} In many real applications such as the data fusion \cite{10.14778/1687553.1687620}, social network
analysis \cite{Sergey2012JointLU}, and the Semantic Web \cite{10.1007/978-3-642-38288-8_24}, one important and fundamental problem is to identify and link the same real-world entities
from multiple data sources, also known as the \textit{entity resolution} (ER) problem \cite{papadakis2020surveyblockingfilteringtechniques}. Specifically, an ER problem retrieves
from data sources the matching pairs of records or profiles that
represent the same entities, which can be inferred by their similar
or the same attribute values. In the era of big data, new data records
often arrive very fast in a streaming fashion, thus, the ER problem
becomes more challenging over dynamic data sources (e.g., data
streams). An efficient solution to such an online ER problem can be
used as a critical step during the process of the data integration.


While many existing works on the ER problem over static data \cite{ebraheem2018distributed} and data streams \cite{Rezig2015-sb,Ren21b} have been widely studied \cite{papadakis2020surveyblockingfilteringtechniques}, they generally assume that the underlying data are complete and accurate. In practice, however, attributes extracted from unstructured text via information extraction (IE) \cite{10.5555/1619645.1619792} are often missing or inconsistent due to unreliable sources. Similarly, on social networks (e.g., Twitter) or health-related forums, users often post daily textual comments/messages such as tweets/retweets (i.e., entities) about different topics/events. Since users may not fully describe their opinions or IE techniques are sometimes not very accurate, some extracted attributes from the unstructured comment/message texts may be missing or even conflicting with each other. In this case, it is rather challenging to conduct online ER operations over such incomplete/inconsistent comment/message streams \cite{Ren21b}.


Graph-structured data has received comparatively little attention \cite{bhattacharya2006entity, Fan22}, and, to best our knowledge, not many previous works have addressed ER over graphs collected from both trusted and untrusted sources.

\vspace{1ex}\noindent{\bf Privacy for Entity Resolution.} Privacy-Preserving Entity Resolution aims to achieve ER while protecting privacy \cite{Gkou21}. Integrating privacy into \textit{reliability-aware entity resolution} (RER) is essential to balance data protection, trust, and match quality. Indeed, ER often involves linking records from multiple datasets, which may contain personally identifiable information (PII) such as names, addresses, financial information, and certain sensitive entity relationships. Privacy mechanisms prevent data leakage when resolving entities between organizations or datasets. RER involves trust scores or confidence levels, which may expose sensitive attributes indirectly. Compliance with privacy laws (e.g. GDPR, and HIPAA) allows us to restrict how personal data can be shared and processed. Moreover, the reliability scores and trust metrics used in RER may disclose sensitive information if adversaries infer the identities of entities based on the trust values.

Adding privacy mechanisms to RER is very challenging. That is, it aims to improve matching accuracy by exploiting uncertainty estimates, confidence scores, or reliability indicators, whereas privacy mechanisms (e.g., differential privacy) introduce noise, obfuscation, or data access restrictions, distorting the reliability signals and making the decision to match two entities more difficult. Therefore, the challenge is to develop algorithms that can handle both inherent uncertainty and privacy-related distortions. Second, many privacy-aware RE methods degrade the quality of similarity scores or block access to certain data, thus reducing the accuracy of matches \cite{Gkou21}. However, RER requires accurate trust indicators, which privacy mechanisms often disrupt or obscure. The challenge, therefore, is to find the optimal tradeoff between privacy and utility, while preserving the efficiency of entity resolution.

Several avenues of research have been explored to carry out private ER \cite{KarapiperisGV17, KhurramK20}. However, all these methods are based on preserving the privacy of sensitive information related to tabular data.

In this paper, we envision the setting that involves applying RER to graph-structured data with uncertainty estimates of different confidence levels, for which the \textit{differential privacy} (DP) approach is a natural fit (i.e., privacy preservation in RER over graphs). DP can be applied either by publishing a sanitized graph \cite{NguyenIR16} or by acting as a proxy that returns optimally noised query answers \cite{Hay2009AccurateEO}, using node DP or edge DP, with node DP able to accommodate node attributes (see the survey in \cite{Yang2023}). Although uncertain graphs were originally used to protect sensitive information by injecting uncertainty into edges \cite{NguyenIR15}, they lack rigorous privacy guarantees, which has motivated combining uncertain graphs with DP \cite{Jing2017}.

Specifically, there are two different existing works of DP to protect sensitive information from an analyst who can query a graph. The first approach is to publish a sanitized graph that is a close view of the original graph, but is ensured to be private, according to DP. The analyst can use this sanitized graph to answer their queries \cite{NguyenIR16}. The second approach is to act as a proxy for the analyst by querying the graph for them. The exact answers to the query can then be optimally noised before being returned to the analyst \cite{Hay2009AccurateEO}. To protect sensitive information in graph data, two forms of DP are used: node DP and edge DP. The node DP could be combined with attributes (e.g. age, gender of some entity) decorating the nodes. For details on how DP is applied to the graph data, see the survey given in \cite{Yang2023}. Although uncertain graphs were originally used to protect sensitive information by injecting uncertainty into the edges of the original graph \cite{NguyenIR15}, their use lacks rigorous privacy guarantees. Therefore, a new research avenue has emerged, combining uncertain graphs and the principles of DP in order to provide formal privacy properties \cite{Jing2017,Lian2023}.

\balance

\section{Conclusion}
There is a fundamental tension between reliability-aware entity ER on attributed graphs and privacy preservation. Reliability information serves both as a key indicator for improving merging accuracy and as a potential vector for amplifying privacy risks. While existing work treats reliability computation and privacy preservation as independent stages of the ER process, this paper argues that their joint design is not only possible but necessary.

We propose a research vision that couples privacy and reliability across three layers: integrating privacy constraints into reliability estimation, allocating privacy budgets to merge decisions, and perturbing reliability metadata before publication. Realizing this vision requires defining reliability-aware privacy semantics, managing privacy–utility trade-offs during computation, defending against reliability-based fingerprinting, ensuring incremental protection in dynamic scenarios, and establishing a unified evaluation framework. The goal is to maximize trustworthy fusion under specified privacy constraints and minimize privacy costs under specified fusion quality requirements, advancing entity resolution (ER) toward a knowledge fusion framework that jointly prioritizes data trustworthiness and user privacy.

\bibliographystyle{ACM-Reference-Format}
\bibliography{anrbib}

@String{Computing = "Computing" }

@String{Computer = "{IEEE} Computer" }

@String{Springer = "Springer-Verlag" }

@inproceedings{Sok26,
author = {D'Silva, Nicholas and Nepal, Surya and Kanhere, Salil S.},
title = {SoK: Practical Aspects of Releasing Differentially Private Graphs},
year = {2026},
isbn = {9798400723568},
publisher = {Association for Computing Machinery},
address = {New York, NY, USA},
url = {https://doi.org/10.1145/3779208.3805972},
doi = {10.1145/3779208.3805972},
booktitle = {Proceedings of the ACM Asia Conference on Computer and Communications Security},
pages = {1799–1815},
numpages = {17},
location = {Bangalore, India},
series = {ASIA CCS '26}
}

@inproceedings{Ren21b,
  author    = {Weilong Ren and
               Xiang Lian and
               Kambiz Ghazinour},
  title     = {Online Topic-Aware Entity Resolution Over Incomplete Data Streams},
  booktitle = {Proceedings of the{ACM} {SIGMOD} International Conference on
               Management of Data},
  pages     = {},
  publisher = {},
  year      = {2015},
  url       = {https://doi.org/10.1145/2723372.2747648},
  doi       = {10.1145/2723372.2747648},
  bibsource = {dblp computer science bibliography, https://dblp.org}
}

@article{ebraheem2018distributed,
  title={Distributed representations of tuples for entity resolution},
  author={Ebraheem, Muhammad and Thirumuruganathan, Saravanan and Joty, Shafiq and Ouzzani, Mourad and Tang, Nan},
  journal={Proceedings of the VLDB Endowment},
  volume={11},
  number={11},
  pages={1454--1467},
  year={2018},
  publisher={VLDB Endowment}
}

@article{10.14778/1687553.1687620,
author = {Dong, Xin Luna and Naumann, Felix},
title = {Data fusion: resolving data conflicts for integration},
year = {2009},
issue_date = {August 2009},
publisher = {VLDB Endowment},
volume = {2},
number = {2},
issn = {2150-8097},
url = {https://doi.org/10.14778/1687553.1687620},
doi = {10.14778/1687553.1687620},
journal = {Proc. VLDB Endow.},
month = aug,
pages = {1654–1655},
numpages = {2}
}

@inproceedings{Sergey2012JointLU,
  title={Joint Link-Attribute User Identity Resolution in Online Social Networks},
  author={Bartunov Sergey and Korshunov Anton and Park Seung-Taek and Ryu Won-Ho and Lee Hyungdong},
  year={2012},
  url={https://api.semanticscholar.org/CorpusID:39334407}
}

@InProceedings{10.1007/978-3-642-38288-8_24,
author="Gangemi, Aldo",
editor="Cimiano, Philipp
and Corcho, Oscar
and Presutti, Valentina
and Hollink, Laura
and Rudolph, Sebastian",
title="A Comparison of Knowledge Extraction Tools for the Semantic Web",
booktitle="The Semantic Web: Semantics and Big Data",
year="2013",
publisher="Springer Berlin Heidelberg",
address="Berlin, Heidelberg",
pages="351--366",
isbn="978-3-642-38288-8"
}

@misc{papadakis2020surveyblockingfilteringtechniques,
      title={A Survey of Blocking and Filtering Techniques for Entity Resolution}, 
      author={George Papadakis and Dimitrios Skoutas and Emmanouil Thanos and Themis Palpanas},
      year={2020},
      eprint={1905.06167},
      archivePrefix={arXiv},
      primaryClass={cs.DB},
      url={https://arxiv.org/abs/1905.06167}, 
}

@INPROCEEDINGS{Rezig2015-sb,
  title           = "Query-time record linkage and fusion over Web databases",
  booktitle       = "2015 {IEEE} 31st International Conference on Data
                     Engineering",
  author          = "Rezig, El Kindi and Dragut, Eduard C and Ouzzani, Mourad
                     and Elmagarmid, Ahmed K",
  publisher       = "IEEE",
  month           =  apr,
  year            =  2015,
  conference      = "2015 IEEE 31st International Conference on Data
                     Engineering (ICDE)",
  location        = "Seoul, South Korea"
}

@inproceedings{10.5555/1619645.1619792,
author = {Poon, Hoifung and Domingos, Pedro},
title = {Joint inference in information extraction},
year = {2007},
isbn = {9781577353232},
publisher = {AAAI Press},
booktitle = {Proceedings of the 22nd National Conference on Artificial Intelligence - Volume 1},
pages = {913–918},
numpages = {6},
location = {Vancouver, British Columbia, Canada},
series = {AAAI'07}
}

@article{bhattacharya2006entity,
  title={Entity resolution in graphs},
  author={Bhattacharya, Indrajit and Getoor, Lise},
  journal={Mining graph data},
  volume={311},
  year={2006},
  publisher={Wiley Hoboken, NJ}
}

@article{Fan22,
author = {Fan, Wenfei},
title = {Big graphs: challenges and opportunities},
year = {2022},
issue_date = {August 2022},
publisher = {VLDB Endowment},
volume = {15},
number = {12},
issn = {2150-8097},
journal = {Proc. VLDB Endow.},
month = aug,
pages = {3782–3797},
numpages = {16}
}

@article{Gkou21,
author={Gkoulalas-Divanis, Aris and Vatsalan, Dinusha and Karapiperis, Dimitrios and Kantarcioglu, Murat},
journal={IEEE Trans on Info Forensics and Security}, 
title={Modern Privacy-Preserving Record Linkage Techniques: An Overview}, 
year={2021},
volume={16},
number={},
pages={4966-4987}}

@inproceedings{KarapiperisGV17,
  author       = {Dimitrios Karapiperis and
                  Aris Gkoulalas{-}Divanis and
                  Vassilios S. Verykios},
  title        = {Distance-Aware Encoding of Numerical Values for Privacy-Preserving
                  Record Linkage},
  booktitle    = {33rd {IEEE} International Conference on Data Engineering, {ICDE} 2017,
                  San Diego, CA, USA, April 19-22, 2017},
  pages        = {135--138},
  publisher    = {{IEEE} Computer Society},
  year         = {2017}
}

@inproceedings{KhurramK20,
  author       = {Basit Khurram and
                  Florian Kerschbaum},
  title        = {SFour: {A} Protocol for Cryptographically Secure Record Linkage at
                  Scale},
  booktitle    = {36th {IEEE} International Conference on Data Engineering, {ICDE} 2020,
                  Dallas, TX, USA, April 20-24, 2020},
  pages        = {277--288},
  publisher    = {{IEEE}},
  year         = {2020}
}

@article{NguyenIR16,
  author       = {Hiep H. Nguyen and
                  Abdessamad Imine and
                  Micha{\"{e}}l Rusinowitch},
  title        = {Network Structure Release under Differential Privacy},
  journal      = {Trans. Data Priv.},
  volume       = {9},
  number       = {3},
  pages        = {215--241},
  year         = {2016}}

@article{Hay2009AccurateEO,
  title={Accurate Estimation of the Degree Distribution of Private Networks},
  author={Michael Hay and Chao Li and Gerome Miklau and David D. Jensen},
  journal={2009 Ninth IEEE International Conference on Data Mining},
  year={2009},
  pages={169-178}
}

@article{Yang2023,
author = {Li, Yang and Purcell, Michael and Rakotoarivelo, Thierry and Smith, David and Ranbaduge, Thilina and Ng, Kee Siong},
title = {Private Graph Data Release: A Survey},
year = {2023},
issue_date = {November 2023},
publisher = {Association for Computing Machinery},
address = {New York, NY, USA},
volume = {55},
number = {11},
issn = {0360-0300},
url = {https://doi.org/10.1145/3569085},
doi = {10.1145/3569085},
journal = {ACM Comput. Surv.},
month = feb,
articleno = {226},
numpages = {39}
}

@inproceedings{NguyenIR15,
  author       = {Hiep H. Nguyen and
                  Abdessamad Imine and
                  Micha{\"{e}}l Rusinowitch},
  editor       = {Feng Bao and
                  Steven Miller and
                  Jianying Zhou and
                  Gail{-}Joon Ahn},
  title        = {Anonymizing Social Graphs via Uncertainty Semantics},
  booktitle    = {Proceedings of the 10th {ACM} Symposium on Information, Computer and
                  Communications Security, {ASIA} {CCS} '15, Singapore, April 14-17,
                  2015},
  pages        = {495--506},
  publisher    = {{ACM}},
  year         = {2015}
}

@INPROCEEDINGS {Jing2017,
author = { Hu, Jing and Shi, Wuchao and Liu, Hai and Yan, Jun and Tian, Yuan and Wu, Zhenqiang },
booktitle = { 2017 International Conference on Networking and Network Applications (NaNA) },
title = {{ Preserving Friendly-Correlations in Uncertain Graphs Using Differential Privacy }},
year = {2017},
volume = {},
ISSN = {},
pages = {24-29},
month =Oct}

@article{Lian2023,
title = {Homogeneous network publishing privacy protection based on differential privacy uncertainty},
journal = {Information Sciences},
volume = {636},
pages = {118925},
year = {2023},
issn = {0020-0255},
author = {Lianwei Qu and Jing Yang and Yong Wang}
}

@article{dwork2014,
  title={The algorithmic foundations of differential privacy},
  author={Dwork, Cynthia and Roth, Aaron and others},
  journal={Foundations and Trends{\textregistered} in Theoretical Computer Science},
  volume={9},
  number={3--4},
  pages={211--407},
  year={2014},
  publisher={Now Publishers, Inc.}
}


\end{document}